\documentclass[sigconf]{acmart}

\AtBeginDocument{%
  \providecommand\BibTeX{{%
    \normalfont B\kern-0.5em{\scshape i\kern-0.25em b}\kern-0.8em\TeX}}}

\copyrightyear{2024}
\acmYear{2024}
\setcopyright{acmlicensed}\acmConference[WWW '24 Companion]{Companion Proceedings of the ACM Web Conference 2024}{May 13--17, 2024}{Singapore, Singapore}
\acmBooktitle{Companion Proceedings of the ACM Web Conference 2024 (WWW '24 Companion), May 13--17, 2024, Singapore, Singapore}
\acmDOI{10.1145/3589335.3648314}
\acmISBN{979-8-4007-0172-6/24/05}

\usepackage{xspace}
\usepackage{color}
\usepackage{xcolor}
\usepackage{makecell}
\usepackage{enumitem}
\usepackage{mdframed}
\usepackage{lipsum}

\usepackage[ruled]{algorithm2e}
\usepackage[normalem]{ulem}
\newcommand{\method}{NoteLLM\xspace}
\usepackage{tablefootnote}
\usepackage{longtable}
\newcounter{myfootnotecounter}

\begin{document}

\title{NoteLLM: A Retrievable Large Language Model for Note Recommendation}

\author{Chao Zhang}
\affiliation{
 \institution{University of Science and Technology of China \& State Key Laboratory of Cognitive Intelligence}
 \city{Hefei}
\country{China}
 }
\email{zclfe00@gmail.com}

 \author{Shiwei Wu}
\affiliation{
 \institution{University of Science and Technology of China \& State Key Laboratory of Cognitive Intelligence}
 \city{Hefei}
\country{China}
 }
\email{dwustc@mail.ustc.edu.cn}

 \author{Haoxin Zhang}
\affiliation{
 \institution{Xiaohongshu Inc.}
 \city{Beijing}
\country{China}
 }
\email{zhanghaoxin@xiaohongshu.com}

 \author{Tong Xu}
 \authornote{Corresponding authors.}
\affiliation{
 \institution{University of Science and Technology of China \& State Key Laboratory of Cognitive Intelligence}
 \city{Hefei}
\country{China}
 }
\email{tongxu@ustc.edu.cn}

 \author{Yan Gao}
\affiliation{
 \institution{Xiaohongshu Inc.}
 \city{Beijing}
\country{China}
 }
\email{yadun@xiaohongshu.com}

 \author{Yao Hu}
\affiliation{
 \institution{Xiaohongshu Inc.}
 \city{Beijing}
\country{China}
 }
\email{xiahou@xiaohongshu.com}

\author{Di Wu}
\affiliation{
 \institution{Xiaohongshu Inc.}
 \city{Beijing}
\country{China}
 }
\email{wudi1123@foxmail.com}

 \author{Enhong Chen}
 \authornotemark[1]
\affiliation{
 \institution{University of Science and Technology of China \& State Key Laboratory of Cognitive Intelligence}
 \city{Hefei}
\country{China}
 }
\email{cheneh@ustc.edu.cn}
\renewcommand{\shortauthors}{Chao Zhang et al.}



\begin{abstract}
People enjoy sharing "notes" including their experiences within online communities. 
Therefore, recommending notes aligned with user interests has become a crucial task.  
Existing online methods only input notes into BERT-based models to generate note embeddings for assessing similarity.
However, they may underutilize some important cues, e.g., hashtags or categories, which represent the key concepts of notes.
Indeed, learning to generate hashtags/categories can potentially enhance note embeddings, both of which compress key note information into limited content.
Besides, Large Language Models (LLMs) have significantly outperformed BERT in understanding natural languages. 
It is promising to introduce LLMs into note recommendation.
In this paper, we propose a novel unified framework called NoteLLM, which leverages LLMs to address the item-to-item (I2I) note recommendation.
Specifically, we utilize Note Compression Prompt to compress a note into a single special token, and further learn the potentially related notes’ embeddings via a contrastive learning approach.
Moreover, we use NoteLLM to summarize the note and generate the hashtag/category automatically through instruction tuning.
Extensive validations on real scenarios demonstrate the effectiveness of our proposed method compared with the online baseline and show major improvements in the recommendation system of Xiaohongshu.

\begin{figure}[!t]
    \centering
    \includegraphics[width=0.48\textwidth]{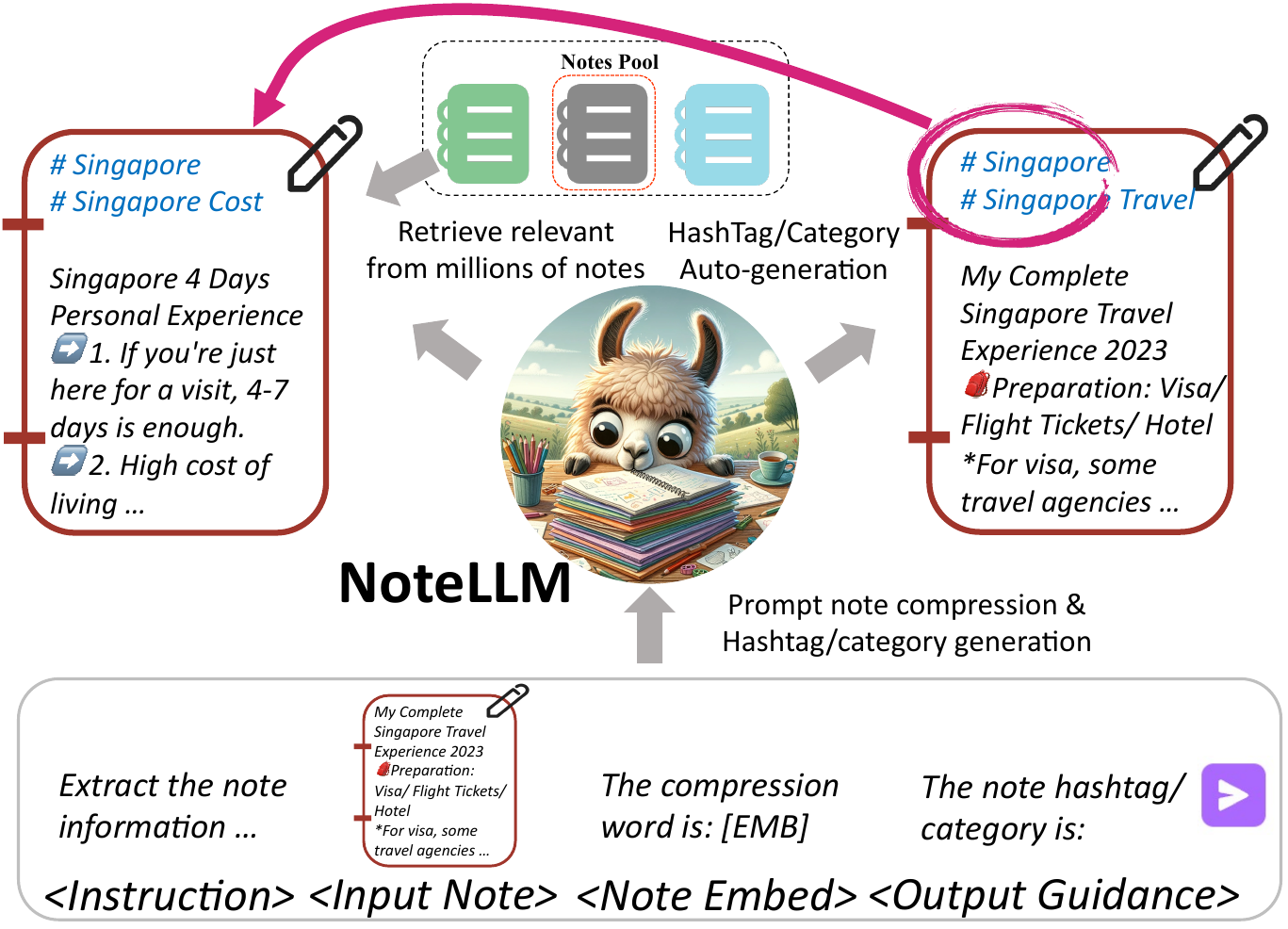}
    \caption{An example of recommending the relevant note from millions-level notes pool via NoteLLM. Learning hashtag generation benefits item-to-item recommendation tasks.}
    
    \label{fig:intro}
\end{figure}


\end{abstract}




\maketitle

\section{Introduction}
    
Focused on user-generated content (UGC) and providing a more authentic and personalized user experience, social media like Xiaohongshu and Lemon8 have gained significant popularity among users. These platforms encourage users to share their product reviews, travel blogs, and life experiences, among other content, also referred to as "notes". 
By providing more personalized notes based on user preferences, note recommendation plays a crucial part in enhancing user engagement~\cite{liao2022user,wu2023personalized,peng2023gpt,zhao2023hierarchical}.
Item-to-item (I2I) note recommendation is a classic way to retrieve notes of potential interest to the user from the millions-level notes pool~\cite{linden2003amazon,zhao2023bootstrapping}.
Given a target note, I2I methods select the relevant notes according to content~\cite{zhao2023bootstrapping} or collaborative signals~\cite{linden2003amazon}.

Existing online methods of I2I note recommendation usually input whole note content into BERT-based models~\cite{devlin2018bert} to generate embeddings of notes, and recommend relevant notes based on embedding similarity~\cite{reimers2019sentence,jiang2022promptbert}.
However, these methods merely treat hashtags/categories as a component of note content, underutilizing their potential.
As shown in Figure~\ref{fig:intro}, hashtags/categories (e.g., \textit{\# Singapore}) represent the central ideas of notes, which are crucial in determining whether two notes contain related content.
In fact, we find that generating hashtags/categories is similar to producing note embeddings.
Both compress the key note information into limited content. 
Therefore, learning to generate hashtags/categories can potentially enhance the quality of embeddings.
Besides, Large Language Models (LLMs) have recently exhibited powerful abilities in natural languages~\cite{xu2023large,jiang2023scaling,lyu2024crud,touvron2023llama2} and recommendations~\cite{peng2023gpt,Bao_2023,bao2023bi,zhang2023recommendation}.
However, there is a scarcity of research investigating the application of LLMs in I2I recommendations. 
Utilizing LLMs to improve I2I note recommendations holds considerable promise.




Inspired by the above insights, we propose a unified multi-task approach called NoteLLM in this paper.
Based on LLMs, NoteLLM learns from the I2I note recommendation and hashtag/category generation tasks, aiming to enhance the I2I note recommendation ability by learning to extract condensed concepts. 
Specifically, we first construct a unified Note Compression Prompt for each note sample and then decode via pre-trained LLMs (e.g., LLaMA 2~\cite{touvron2023llama2}), which utilize a special token to compress the note content and generate hashtags/categories simultaneously.
To construct the related note pairs, we count the co-occurrence scores for all note pairs from user behaviours, and form the set of co-occurrence scores for each note.
We select notes with the highest co-occurrence scores in the set as the related notes for a given note.
Further, to recommend the relevant notes for each sample, Generative-Contrastive Learning (GCL) utilizes the compressed tokens as the embedding of each note, and then trains the LLMs to identify the related notes from in-batch negatives.
Simultaneously, we employ Collaborative Supervised Fine-tuning (CSFT) approach to train models to generate hashtags/categories for each note. 
Since both the compression token learned by the I2I note recommendation task and the hashtag/category generation task aim to extract the key concept of the note content, CSFT can enhance note embeddings effectively.

Our paper makes the following contributions:
\begin{itemize}[leftmargin=*]
    \item To the best of our knowledge, our NoteLLM framework is the first to address the I2I recommendation task utilizing LLMs.
    It reveals that introducing LLMs is a practical and promising strategy to enhance I2I recommendation systems.
    \item We propose a multi-task framework to learn I2I recommendation task and hashtag/category generation task to enhance note embeddings.
    We demonstrate that learning to generate the compressed concepts is beneficial to the I2I recommendation task.
    \item Extensive validations on offline experiments and online industrial scenarios of Xiaohongshu demonstrate the effectiveness of our proposed technical framework for note recommendation.
\end{itemize}

\section{Related work}
\subsection{I2I Recommendation}
I2I recommendation is a crucial technique that can recommend a ranked list of items from a large-scale item pool based on a target item.
I2I recommendation either pre-constructs the I2I index~\cite{yang2020large} or retrieves relevant items online using the approximate k-nearest neighbor method~\cite{johnson2019billion}.
Traditional I2I recommendations typically rely solely on collaborative signals from user behaviors~\cite{zhu2018learning,yang2020large}.
However, these methods cannot manage cold-start items due to lack of user-item interaction~\cite{zhao2023bootstrapping}.
To address this issue, numerous studies have investigated content-based I2I recommendations~\cite{huang2013learning,zhao2023bootstrapping}.
We focus on the text-based I2I recommendation system, which measures the similarity of items based on their textual content.
Initially, representation of text-based I2I recommendation relied on a term-based sparse vector matching mechanism~\cite{robertson2009probabilistic,ramos2003using}.
With the advent of deep learning, neural networks have proven more adept at representing text information~\cite{mikolov2013efficient,devlin2018bert}.
Previous works~\cite{karpukhin2020dense, ma2022pre, xiao2022training, xiao2022progressively} transform texts into embeddings in the same latent space to measure their relationship through embedding similarity.
LLMs have recently gained great attention for their impressive abilities~\cite{xu2023large,yin2023survey,peng2023large}.
However, the application of LLMs in I2I recommendation remains unexplored.
Besides, some studies treat LLMs solely as encoders for generating embeddings~\cite{jiang2023scaling,muennighoff2022sgpt,ma2023fine}, failing to leverage their full potential for generation.
In \method, we utilize LLMs to generate hashtags/categories, which can enhance note embeddings.

\subsection{LLMs for Recommendation}
LLMs have recently made significant advancements~\cite{touvron2023llama,touvron2023llama2,openai2023gpt4}.
Consequently, numerous studies incorporate LLMs into recommendation tasks~\cite{wu2023survey,lin2023can,fan2023recommender}.
There are three main methods of integrating LLMs with recommendations~\cite{wu2023survey,lin2023can}.
The first method is utilizing LLMs to augment data~\cite{liu2023first,xi2023towards,mysore2023large}.
Due to the abundant world knowledge contained by LLMs, the augmented data are more prominent and diverse than the raw data~\cite{xi2023towards,wang2023anypredict,lyu2023llm}.
However, these methods require continuous preprocessing of the testing data to align with the augmented training data and are highly dependent on the quality of LLMs' generation.
The second method is leveraging LLMs to recommend directly.
These methods design special prompts~\cite{wang2023recmind,huang2023recommender,liu2023chatgpt} or use supervised finetuning~\cite{Bao_2023,bao2023bi,zhang2023recommendation} to induce LLMs to answer the given questions. 
Nevertheless, because of the limited context length, these methods only focus on the reranking stage~\cite{hou2023large,zhang2023recommendation}, which only contains dozens of candidate items.
The last method is adopting LLMs as the encoders to generate embeddings representing specific items~\cite{li2023exploring,wu2021empowering}.
Although these methods are effective to extract information, they all discard the generative capabilities of LLMs.
In contrast to above methods, \method employs LLMs during the recall phase and learns hashtag generation to improve LLMs' ability to produce embeddings.

\subsection{Hashtag/Category Generation from Text}
Hashtags and categories, as tagging mechanisms on social media, streamline the identification of topic-specific messages and aid users in finding themed content. Generating these from text can assist in creating identifiers for untagged notes or suggesting options to users based on their preferences.
In this domain, there are three main methods: extractive, classification, and generative methods.
Extractive methods identify key phrases in texts as hashtags or categories~\cite{zhang2016keyphrase,zhang2018encoding}, but cannot obtain those not present in the original text.
Classification methods view this task as a text classification problem~\cite{zeng2018topic,zhang2017hashtag,kou2018hashtag}.
However, these may yield sub-optimal results due to the diverse, free-form nature of human-generated hashtags.
Generative methods generate the hashtags/categories directly according to input texts~\cite{wang2019microblog,wang2019topic,diao2023hashtag}.
Whereas, these methods are limited to solving the hashtag/category generation task.
In \method, LLMs perform multi-task learning, simultaneously executing I2I recommendation and hashtag/category generation.
Due to the similarity of these two tasks, learning to generate the hashtag/category can also enhance the I2I recommendation.


\section{Problem Definition}


In this section, we introduce the problem definition.
We assume $\mathcal{N}=\{n_1,n_2,...,n_m\}$ as note pool, where $m$ is the number of notes.
Each note contains a title, hashtag, category, and content. 
We denote the $i$-th note as $n_i=(t_i, tp_i, c_i,ct_i)$, where $t_i$, $tp_i$, $c_i$, $ct_i$ mean the title, the hashtag, the category and the content respectively.
In the I2I note recommendation task, given a target note $n_i$, the LLM-based retriever aims to rank the top-$k$ notes, which are similar to the given note, from the note pool $\mathcal{N} \backslash \{n_i\}$.
In the hashtag/category generation task, the LLM is utilized to generate the hashtag $tp_i$ according to $t_i$ and $ct_i$.
Besides, in the category generation task, the LLM is to generate the category $c_i$ according to $t_i$, $tp_i$ and $ct_i$.

\section{Methodology}
\begin{figure*}[!h]
    \centering
    \includegraphics[width=\textwidth]{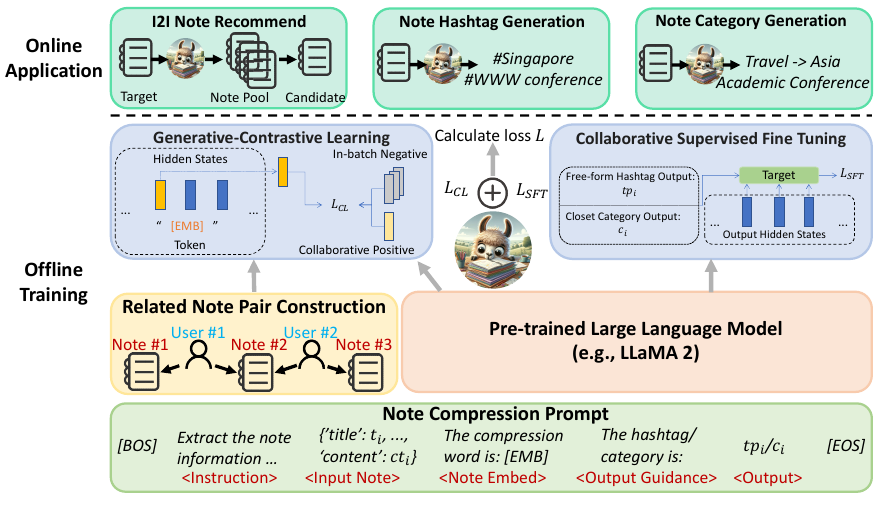}
    \caption{The \method framework uses a unified prompt for I2I note recommendations and hashtag/category generation. Notes are compressed via the Note Compression Prompt and processed by pre-trained LLMs. We utilize the co-occurrence mechanism to construct the related note pairs and train the I2I recommendation task using Generative-Contrasting Learning. \method also extracts note's key concepts for hashtag/category generation, enhancing the I2I recommendation task.}
    \label{fig:framework}
\end{figure*}

\subsection{Framework of \method}
In this subsection, we introduce the framework of \method, which comprises three key components: Note Compression Prompt Construction, GCL, and CSFT, as illustrated in Figure~\ref{fig:framework}.
We employ Note Compression Prompt to flexibly manage the I2I recommendation and hashtag/category generation tasks.
These prompts are then tokenized and fed into LLMs.
\method integrates both collaborative signals and semantic information into the hidden states. GCL uses the hidden states of the generated compressed word to conduct contrastive learning, thereby acquiring collaborative signals.
Furthermore, CSFT leverages the semantic and collaborative information of the note to generate hashtags and categories.


\subsection{Note Compression Prompt}
We employ a unified Note Compression Prompt to facilitate both I2I recommendation and generation tasks. To leverage the generative capabilities of autoregressive LLMs for I2I recommendation tasks~\cite{jiang2023scaling}, our aim is to condense the note content into a single special token. This condensed special token is then used to acquire collaborative knowledge through GCL. Subsequently, we generate hashtags/categories using this knowledge via CSFT.

Specifically, we propose the following prompt template for general note compression and hashtags/categories generation:

\begin{mdframed}
    \textbf{Prompt:} [BOS]<Instruction> <Input Note> The compression word is:"[EMB]". <Output Guidance> <Output>[EOS]
\end{mdframed}

In this template, [BOS], [EMB], and [EOS] are special tokens, while <Instruction>, <Input Note>, <Output Guidance>, and <Output> are placeholders replaced by specific content. The specific content for category generation is defined as follows:

\begin{mdframed}
    \textbf{Note Compression Prompt for Category Generation.} \\
    \textbf{<Instruction>:} Extract the note information in json format, compress it into one word for recommendation, and generate the category of the note. \\
    \textbf{<Input Note>:} \{'title': $t_i$ , 'topic': $tp_i$ , 'content': $ct_i$\}. \\
    \textbf{<Output Guidance>:} The category is: \\
    \textbf{<Output>:} $c_i$
\end{mdframed}

The template for hashtag generation is presented below:
\begin{mdframed}
    \textbf{Note Compression Prompt for Hashtag Generation.} \\
    \textbf{<Instruction>:} Extract the note information in json format, compress it into one word for recommendation, and generate <j> topics of the note. \\
    \textbf{<Input Note>:} \{'title': $t_i$ , 'content': $ct_i$\}. \\
    \textbf{<Output Guidance>:} The <j> topics are: \\
    \textbf{<Output>:} <j> topics from $tp_i$
\end{mdframed}
Given the unpredictability of the number of hashtags generated by users, we randomly select a subset of original hashtags as the output target for hashtag generation to minimize potential misguidance to LLMs. The number of randomly selected hashtags, denoted as <j>, is incorporated into both the <Instruction> and <Output Guidance>.

Once the prompts are completed, they are tokenized and fed into the LLM. The LLM then distills the collaborative signals and key semantic information into the compressed word and generates hashtags/categories based on the central ideas of notes.


\subsection{Generative-Contrastive Learning}
Pre-trained LLMs usually learn new knowledge via instruction tuning~\cite{zhang2023instruction,wei2021finetuned} or Reinforcement Learning from Human Feedback (RLHF)~\cite{stiennon2020learning,ouyang2022training}.
These methods mainly focus on leveraging semantic information to enhance the effectiveness and safety of the LLMs.
However, relying solely on semantic information in LLMs is insufficient for recommendation tasks~\cite{liu2023chatgpt,he2023large}. 
Collaborative signals, which are absent in LLMs, play a vital role in identifying the notes that are of specific interest to users~\cite{liu2023chatgpt,he2023large}.
Therefore, we propose GCL to empower LLMs to capture stronger collaborative signals.
In contrast to learning from specific answers or reward models, GCL adopts contrastive learning, which learns the relational proximity among notes from a holistic perspective.

In order to integrate collaborative signals into LLMs, we adopt the co-occurrence mechanism to construct the related note pairs based on user behaviours.
This mechanism is based on the assumption that notes frequently read together are likely related.
Therefore, we collect user behavior data within one week for the co-occurrence count.
We count the occurrences in which users viewed note $n_A$ and subsequently clicked on note $n_B$.
Simultaneously, to differentiate the contribution of co-occurrence from different users, we assigned varying weights to distinct clicks.
Specifically, we compute the co-occurrence score as following:
\begin{equation}
s_{n_A\rightarrow n_B} = \sum^{U}_{i=1}\frac{1}{N_i},
\end{equation}
where $s_{n_A\rightarrow n_B}$ represents the co-occurrence score from note $n_A$ to note $n_B$, $U$ is the number of users in this user behavior data, and $N_i$ denotes the quantity of the note set clicked by the $i$-th user in the user behavior data.
This operation aims to prevent the misdirection of active users, who might indiscriminately click on every note recommended to them.
After calculating the co-occurrence score for all note pairs,
we construct the set of co-occurrence scores $\mathcal{S}_{n_i}$ from note $n_i$ to all other notes. Specifically, $\mathcal{S}_{n_i}$ is defined as $\{{s_{n_i\rightarrow n_j}}|1\leq j \leq m, i\neq j\}$.
Next, we filter outlier notes whose co-occurrence scores are either above $u$ or below the threshold $l$ from $\mathcal{S}_{n_i}$.
Finally, we select the $t$ notes with the highest co-occurrence scores from the filtered set as the related notes for note $n_i$.

After constructing the related notes pairs, we train \method to determine the relevance of notes based on textual semantics and collaborative signals.
Different from simply taking a special pooling word to represent the note~\cite{ma2023fine}, we utilize prompts to compress the note information to generate one virtual word.
The last hidden state of the compressed virtual word contains the semantic information and collaborative signals of the given note, which can represent the note.
Specifically, due to the autoregressive nature of LLMs, we take the last hidden state of the previous token of [EMB] and use a linear layer to transform it to note embedding space, whose dimension is $d$.
We denote the embedding of $i$-th note $n_i$ as $\boldsymbol{n}_i$.
We assume each minibatch contains $B$ related note pairs, resulting in a total of $2B$ notes per minibatch.
We denote the related note of the note $n_i$ as $n_i^+$, and its embedding as $\boldsymbol{n}_i^+$.
Following~\cite{neelakantan2022text}, the loss of GCL is computed as follows:
\begin{equation}
L_{cl} =- \frac{1}{2B}\sum_{i=1}^{2B}log\frac{e^{sim(\boldsymbol{n}_i,\boldsymbol{n}_i^+)\cdot e^{\tau}}}{\sum_{j\in [2B]\backslash\{i\}}e^{sim(\boldsymbol{n}_i,\boldsymbol{n}_j)\cdot e^{\tau}}},
\end{equation}
where $L_{cl}$ denotes the loss of GCL, $\tau$ means the learnable temperature and $sim(a, b) = a^\top b /(\Vert a \Vert \Vert b \Vert)$.


\subsection{Collaborative Supervised Fine-Tuning}

LLMs have gained prominence due to their robust capabilities in semantic understanding and generation.
Several existing works attempt to apply the impressive abilities of LLMs to sentence embeddings~\cite{jiang2023scaling,ma2023fine,muennighoff2022sgpt,su2022one,neelakantan2022text}.
However, these methods overlook the generative capabilities of LLMs, reducing them to mere embedding generators and failing to fully exploit their potential.
Besides, these methods underutilize hashtags/categories, which represent the key concepts of notes.
In fact, generating hashtags/categories is similar to producing note embeddings. 
Both tasks aim to summarize note content.
The task of generating hashtags/categories extracts key note information from a text generation perspective, while the task of producing note embeddings compresses notes into a virtual word from a collaborative viewpoint for I2I recommendation.
To this end, our NoteLLM jointly models the GCL and CSFT tasks to potentially enhance the quality of embeddings.
We integrate these two tasks into a single prompt, providing additional information for both tasks and streamlining the training process.



Specifically, we adopt CSFT, which leverages the semantic content of the notes and the collaborative signals in the compressed token to generate hashtags/categories. 
To enhance training efficiency and prevent the forgetting problem~\cite{toneva2018empirical}, we select $r$ notes from each batch for the hashtag generation task, while the remaining notes are allocated for the category generation task.
Specifically, we compute the CSFT loss as follows:
\begin{equation}
L_{gen}=-\frac{1}{T}\sum_{i=1}^{T}log(p(o_i|o_{<i},i)),
\end{equation}
where $L_{gen}$ is the CSFT loss, $T$ is the length of the output, $o_i$ means the $i$-th token in output sequence $o$ and $i$ is the input sequence. 

Finally, we define the loss function of \method to incorporate both GCL and CSFT, as follows:
\begin{equation}
    L=\frac{L_{cl}+\alpha L_{gen}}{1+\alpha},
    \label{eq:total_loss}
\end{equation}
where $L$ is the total loss of \method and $\alpha$ is the hyperparameter.
Through model updates, NoteLLM is capable of concurrently executing I2I recommendation tasks and hashtag/category generation tasks for note recommendation scenarios.

\section{Experiments}

\subsection{Dataset and Experiment Setting}

    
    

\begin{table}[!h]
    \centering
    
    \setlength\tabcolsep{0.6mm}
    \renewcommand\arraystretch{0.93}
    \caption{Detailed statistics of training and testing dataset.}
    \begin{tabular}{l|r|l|r} \Xhline{1.0pt}
    \multicolumn{4}{c}{training dataset} \\
    \Xhline{1.0pt}
    \#notes & 458,221 & \#note pairs & 312,564 \\

    avg. \#words per title & 11.54 & avg. \#hashtag per note & 3.02\\

    avg. \#words per hashtag & 4.19 & avg. \#words per content & 47.67 \\
    
    \Xhline{1.0pt}
    \multicolumn{4}{c}{testing dataset} \\
    \Xhline{1.0pt}
    \#notes & 257,937 & \#note pairs & 27,999 \\

    avg. \#words per title & 13.70 & avg. \#hashtag per note & 5.49\\

    avg. \#words per hashtag & 4.53 & avg. \#words per content & 182.45 \\
    
    \Xhline{1.0pt}
    \end{tabular}
    \label{tab:testingdatasets}
\end{table}

We conduct offline experiments on Xiaohongshu product datasets. 
To balance the model's training, we generate the training set by extracting a fixed number of note pairs from each category combination based on a week's product data.
Then, we randomly select notes from the upcoming month to form the note pool of testing set, excluding any notes that are already in the training dataset.
The detailed statistics of the training and testing dataset are shown in Table~\ref{tab:testingdatasets}.
Besides, there are more than $500$ categories in our dataset.

In our experiments, we leverage Meta LLaMA 2~\cite{touvron2023llama2} as the base LLMs. 
In related note pair construction, we set the upper bound of the co-occurrence score $u$ as $30$ and the lower bound $l$ as $0.01$.
And we set $t$ as $10$.
Besides, the dimension $d$ of note embedding is set to $128$. 
The batch size $B$ is set to $64$\footnote{We utilize Distributed Data Parallel training on 8 $\times$ 80GB Nvidia A100 GPUs and the
batch size per each GPU is $8$.}.
Each batch contains $128$ notes.
Due to context length restriction, we truncate the titles exceeding $20$ tokens, and truncate the contents exceeding $80$ tokens.
The temperature $\tau$ is initialized as $3$.
We set $\alpha$ in Equation~\ref{eq:total_loss} to $0.01$.
The ratio $r$ for the hashtag generation task is set at $40\%$.

To assess the offline performance of the I2I recommendation model, we choose the prompt for category generation, which contains all input note information. We select the first note from each note pair as the target note, and the other as the ground truth.
Subsequently, we rank all the notes in the test pool, excluding the target note, according to the target note.
We then use Recall@100, Recall@1k, Recall@10k and Recall@100k to validate the model effectiveness for I2I note recommendation. 
For closet category generation tasks, we use accuracy (Acc.) and illusory proportions (Ill.) as the evaluation metrics.
Ill. represent the proportion of categories generated by the model that are not in the closet.
For free-form hashtag generation tasks, we use BLEU4, ROUGE1, ROUGE2 and ROUGEL to evaluate models.

\subsection{Offline Performance Evaluation}
In this subsection, we demonstrate the effectiveness of \method for I2I note recommendation.
We compare our \method with the following text-based I2I recommendation methods:
\begin{itemize}[leftmargin=*]
\item \textbf{zero-shot} utilizes LLMs to generate the embeddings without any prompts and then conducts zero-shot retrieval.
\item \textbf{PromptEOL zero-shot}~\cite{jiang2023scaling} is a zero-shot LLMs sentence embedding method that uses the explicit one-word limitation prompt.
\item \textbf{SentenceBERT}~\cite{reimers2019sentence} adopts BERT to learn the note similarity based on contrastive learning, serving as the online baseline.
\item \textbf{PromptEOL+CSE}~\cite{jiang2023scaling} uses the explicit one-word limitation prompt and leverages contrastive learning to update LLMs.
\item \textbf{RepLLaMA}~\cite{ma2023fine}, a bi-encoder dense retriever based on LLMs without any prompts.
\end{itemize}

\begin{table*}[!h]
    \centering
    
    \setlength\tabcolsep{5pt}
    \renewcommand\arraystretch{1.2}
    \caption{Performance of different methods in I2I recommendation tasks (\%).}
    \begin{tabular}{l|c|ccccc} 
    \Xhline{1.0pt}
    &Model Size&Recall@100&Recall@1k&Recall@10k&Recall@100k & Avg. \\
    \Xhline{1.0pt}
    LLaMA 2 zero-shot & 7B & 11.94 & 19.44 & 32.53 & 68.81 & 33.18 \\
    PromptEOL zero-shot~\cite{jiang2023scaling} & 7B & 55.27 & 74.47 &88.71 & 98.04 & 79.12 \\
    \Xhline{0.7pt}
    SentenceBERT (Online)~\cite{reimers2019sentence} & 110M & 70.72 & 87.88 & 96.29 & 99.62 & 88.63 \\
    PromptEOL+CSE~\cite{jiang2023scaling} & 7B & 83.28 & \textbf{95.26} &99.20 & \textbf{99.96} & 94.43 \\
    RepLLaMA~\cite{ma2023fine} &7B & 83.63 & 95.10 & \textbf{99.27} & 99.94 & 94.49 \\
    \method &7B & \textbf{84.02} & 95.23 & 99.23 & \textbf{99.96} & \textbf{94.66} \\
    
    \Xhline{1.0pt}
    \end{tabular}
    \label{tab:retrieval}
\end{table*}

\begin{table*}[!h]
    \centering
    
    \setlength\tabcolsep{5pt}
    \renewcommand\arraystretch{1.2}
    \caption{Performance of different methods for low exposure notes and high exposure notes in I2I recommendation tasks (\%).}
    \begin{tabular}{l|cc|cc|cc} 
    \Xhline{1.0pt}
     & \multicolumn{2}{c|}{Low Exposure} & \multicolumn{2}{c|}{High Exposure} & \multicolumn{2}{c}{Overall} \\
     \Xhline{1.0pt}
    &Recall@100&Recall@1k&Recall@100&Recall@1k & Recall@100&Recall@1k \\
    \Xhline{1.0pt}
    SentenceBERT (Online)~\cite{reimers2019sentence} & 75.00 & 90.54 & 59.03 & 81.91 & 70.72 & 87.88 \\
    PromptEOL+CSE~\cite{jiang2023scaling} & 86.28 & \textbf{96.63} & 72.46 & \textbf{91.40} & 83.28 & \textbf{95.26} \\
    RepLLaMA~\cite{ma2023fine} & 86.54 & 96.18 & 72.64 & 91.37 & 83.63 & 95.10 \\
    NoteLLM & \textbf{87.85} & \textbf{96.63} & \textbf{73.46} & 91.26 & \textbf{84.02} & 95.23 \\

    \Xhline{1.0pt}
    \end{tabular}
    \label{tab:exposure}
\end{table*}

The results, as presented in Table~\ref{tab:retrieval}, offer several insightful observations.
Despite their potential, zero-shot methods are still unable to surpass the performance of fine-tuned methods, suggesting that the latter's specific knowledge in the note recommendation domain gives them an edge.
Further, we find that the comparison between methods based on LLaMA 2 and SentenceBERT reveals a significant advantage for the former, indicating a superior ability of LLMs to understand notes.
The performance of PromptEOL+CSE with specific prompts matches that of RepLLaMA without prompts, indicating that prompts boost zero-shot retrieval but their effect lessens after fine-tuning.
Lastly, our \method outperforms other LLM-based methods, primarily due to CSFT's effective transfer of summary ability into note embedding compression, which efficiently distills key points for improved note embeddings.



\subsection{Effect on Different Exposure Notes}

In this subsection, we demonstrate the efficacy of our \method in handling notes with varying levels of exposure. For a more comprehensive analysis, we have divided the ground truth notes into two distinct categories based on their exposure levels.
The first category encompasses notes with low-exposure, specifically those with an exposure of less than $1,500$. Despite constituting $30\%$ of all test notes, their cumulative exposure only amounts to $0.5\%$.
On the other hand, the second category includes notes with high-exposure, characterized by an exposure exceeding $75,000$. Even though they represent only $10\%$ of all test notes, their collective exposure is substantial, accounting for $75\%$ of the total.
We then separately calculate the recall for these two groups to further understand the performance of our \method across different exposure levels.


The results are presented in Table~\ref{tab:exposure}. \method consistently outperforms other methods for both low and high exposure notes in most cases, which indicates that the incorporation of CSFT module provides consistent benefits across all notes, irrespective of their exposure levels.
It's worth noting that while these methods exhibit commendable performance with low-exposure notes, they falter when dealing with high-exposure notes. The decline in performance can be attributed to neglecting the popularity bias~\cite{lin2022quantifying}.
Such properties enhance the model's ability to recall based on the content of the notes, making it particularly suitable for retrieving cold-start notes.
This can motivate users to post more new notes, creating richer content for the entire community.


\subsection{Ablation Study}
In this subsection, we conduct an ablation study to underscore the effectiveness of the key innovations in our work. To enhance our analysis, we also show performance on the category and hashtag generation tasks in the following experiments.

\begin{table*}[!h]
    \centering
    
    \setlength\tabcolsep{5pt}
    \renewcommand\arraystretch{1.2}
    \caption{Ablation study for \method in I2I recommendation, category generation and hashtag generation tasks (\%). }
    \begin{tabular}{c|ccccc|cc|cccc} 
    \Xhline{1.0pt}
    Model&R@100&R@1k&R@10k&R@100k & Avg. & Acc. & Ill. & BLEU4 & ROUGE1 & ROUGE2 & ROUGEL \\
    \Xhline{1.0pt}
    NoteLLM & \textbf{84.02} & 95.23 & \textbf{99.23} & \textbf{99.96} & \textbf{94.66} & 66.17 & 0.50 & \textbf{1.38} & 22.31 & 8.02 & 21.03 \\
    w/o CSFT& 83.28 & \textbf{95.26} & 99.20 & \textbf{99.96} & 94.43 & 0.00 & 100.00 & 0.01 & 0.00 & 0.00 & 0.00 \\
    w/o GCL ($r=40\%$) & 75.38 & 90.33 & 97.09 & 98.93 & 90.43 & 75.12 & 2.27 & 1.28 & 26.50 & 13.02 & 23.27 \\
    w/o GCL ($r=0\%$) & 60.38 & 83.22 & 96.13 & 98.84 & 84.64 & \textbf{80.64} & \textbf{0.07} & 0.00 & 1.54 & 0.00 & 1.54 \\
    w/o GCL ($r=100\%$) & 71.98 & 87.86 & 95.59 & 98.51 & 88.49 & 0.18 & 99.70 & 1.30 & \textbf{27.66} & \textbf{14.19} & \textbf{24.11} \\
    
    \Xhline{1.0pt}
    \end{tabular}
    \label{tab:ablation}
\end{table*}

\begin{table*}[!h]
    \centering
    
    \setlength\tabcolsep{5pt}
    \renewcommand\arraystretch{1.2}
    \caption{Performance of \method under different data diversity in CSFT module for I2I recommendation, category generation and hashtag generation tasks (\%).}
    \begin{tabular}{c|ccccc|cc|cccc} 
    \Xhline{1.0pt}
    $r$&R@100&R@1k&R@10k&R@100k & Avg. & Acc. & Ill. & BLEU4 & ROUGE1 & ROUGE2 & ROUGEL \\
    \Xhline{1.0pt}
    0\%& 83.29 & 95.07 & 99.14 & 99.96 & 94.37 & \textbf{71.00} & 0.12 & 0.00 & 0.00 & 0.00 & 0.00 \\
    20\% & 83.81 & \textbf{95.28} &99.26 & 99.96 & 94.58 & 69.70 & \textbf{0.09} & \textbf{1.59} & \textbf{22.57} & 7.73 & \textbf{21.64} \\
    40\% & \textbf{84.02} & 95.23 & 99.23 & 99.96 & \textbf{94.66} & 66.17 & 0.50 & 1.38 & 22.31 & 8.02 & 21.03 \\
    60\% & 83.37 & 95.03 & 99.25 & 99.96 & 94.40 & 63.37 & 0.90 & 1.33 & 21.92 & 7.88 & 20.62 \\
    80\% & 83.15 & 95.06 & \textbf{99.27} & \textbf{99.97} & 94.36 & 53.60 & 2.67 & 1.34 & 22.48 & \textbf{8.31} & 21.10 \\
    100\% & 82.49 & 94.49 & 99.11 & 99.96 & 94.01 & 0.00 & 100.00 & 1.33 & 21.88 & 7.99 & 20.47 \\

    \Xhline{1.0pt}
    \end{tabular}
    \label{tab:diversity}
\end{table*}

We compare our \method with following variants:
\begin{itemize}[leftmargin=*]
\item \textbf{w/o CSFT}, a method solely employs the GCL module.
\item \textbf{w/o GCL ($r=40\%$)} only adopts CSFT module to guide LLMs in summarizing the hashtags and categories.
\item \textbf{w/o GCL ($r=0\%$)} only has category summary task.
\item \textbf{w/o GCL ($r=100\%$)} only instructs LLMs to summarize hashtags.
\end{itemize}

\begin{table*}[!h]
    \centering
    
    \setlength\tabcolsep{5pt}
    \renewcommand\arraystretch{1.2}
    \caption{Performance of \method under varying CSFT module magnitudes for I2I recommendation, category generation and hashtag generation tasks (\%).}
    \begin{tabular}{c|ccccc|cc|cccc} 
    \Xhline{1.0pt}
    Model&R@100&R@1k&R@10k&R@100k & Avg. & Acc. & Ill. & BLEU4 & ROUGE1 & ROUGE2 & ROUGEL \\
    \Xhline{1.0pt}
    $\alpha=0$& 83.42 & 95.13 &\textbf{99.31} & 99.96 & 94.46 & 0.00 & 100.00 & 0.05 & 0.91 & 0.01 & 0.87 \\
    $\alpha=0.001$& 83.73 & 95.09 &99.26 & \textbf{99.97} & 94.51 & 49.04 & 6.84 & \textbf{1.82} & 15.13 & 3.23 & 14.80 \\
    $\alpha=0.01$& \textbf{84.02} & \textbf{95.23} & 99.23 & 99.96 & \textbf{94.66} & 66.17 & 0.50 & 1.38 & 22.31 & 8.02 & 21.03 \\
    $\alpha=0.1$& 83.51 & 95.18 & \textbf{99.31} & 99.96 & 94.49 & 73.33 & \textbf{0.04} & 1.33 & 24.44 & 10.42 & 22.28 \\
    $\alpha=1$& 83.79 & 94.79 & 99.26 & \textbf{99.97} & 94.45 & 74.69 & 0.82 & 1.33 & 26.94 & 13.21 & 23.90 \\
    $\alpha=10$& 82.92 & 94.28 & 98.94 & 99.96 & 94.03 & \textbf{75.15} & 2.08 & 1.30 & \textbf{27.63} & \textbf{13.95} & \textbf{24.21} \\

    \Xhline{1.0pt}
    \end{tabular}
    \label{tab:CSFT}
\end{table*}

The results are presented in Table~\ref{tab:ablation}, from which we can draw several conclusions. Firstly, we observe that the ablation without CSFT performs worse than \method in I2I recommendation task, and completely loses the ability to generate hashtags and categories. This highlights the crucial role of the CSFT module in enhancing note embeddings and suggests that a single model can handle both recommendation and generation tasks.
Secondly, we find that the ablation without GCL module outperforms PromptEOL zero-shot in I2I recommendation tasks. This indicates that the hashtag/category generation task can enhance I2I recommendation tasks. 
Thirdly, the ablation without GCL ($r=40\%$) performs better in I2I recommendation tasks than the version without GCL ($r=0\%$) and without GCL ($r=100\%$). This suggests that task diversity is important for CSFT~\cite{longpre2023flan}.
Finally, we observe a clear seesaw phenomenon~\cite{zheng2023gpt} for hashtag and category generation tasks. The ablation without GCL ($r=0\%$) can generate correct categories for $80.64\%$ of notes, but struggles to summarize useful hashtags. Conversely, the version without GCL ($r=100\%$) generates high-quality hashtags, but the generated categories are mostly incorrect.


\subsection{Impact of Data Diversity in CSFT Module}

In this subsection, we investigate the impact of data diversity in CSFT module.
The performance of the model for different tasks under varying data type proportions is presented in Table~\ref{tab:diversity}.

For I2I recommendation tasks, performance improves as $r$ increases. This is attributed to the enhanced data diversity of instruction tuning with a higher $r$, which more effectively instructs LLMs to summarize and compress various types of information. However, as $r$ continues to increase and the data for instruction tuning becomes more skewed towards the hashtag generation task, performance begins to decline.
For category generation tasks, as the data becomes more biased towards the hashtag generation task, the performance on category generation deteriorates.
However, as $r$ continues to increase from $20\%$, there is not much significant change in the hashtag generation task.
This may be because the category generation task is a closet task, which requires a stringent match.
In contrast, the hashtag generation task is a free-form generation task, which allows for greater flexibility.



\subsection{Impact of the Magnitude of CSFT Module}
In this subsection, we explore the impact of the magnitude of CSFT module on task performance. 
We present the results of our experiments in Table~\ref{tab:CSFT}. Our findings suggest that a slight increase in $\alpha$ enhances the performance of both recommendation and generation tasks. However, as $\alpha$ continues to increase, the performance of the recommendation task begins to decline, while the performance of the generation task continues to improve. This reveals a trade-off between generation and I2I recommendation tasks, highlighting the need for a balanced approach.




\subsection{Case Study}
\begin{figure*}[!h]
    \centering
    \includegraphics[width=0.95\textwidth]{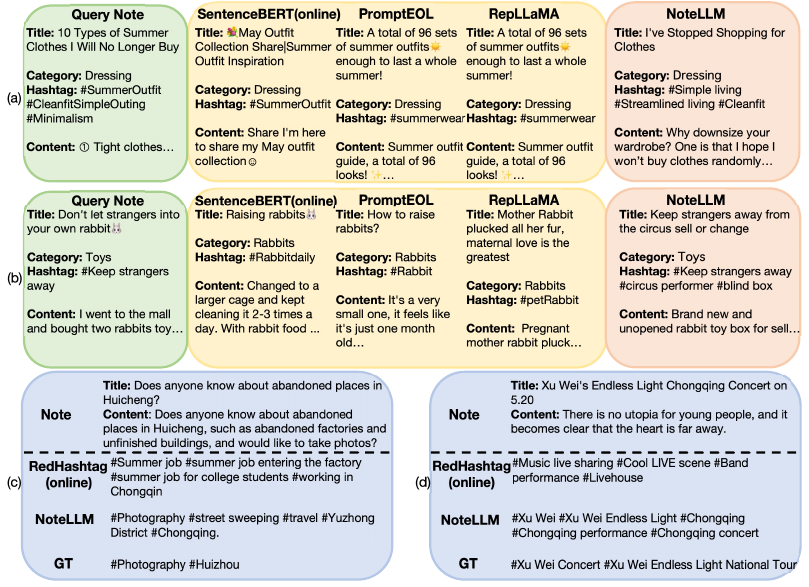}
    \caption{The visualization cases of \method and other baselines. Figure 3(a) and 3(b) show the cases in I2I recommendation tasks, where the left query note is the user's clicked note, and the remaining notes are the top-$1$ ranked results retrieved by different methods. Figure 3(c) and 3(d) show the cases in hashtag generation tasks. RedHashtag is the online hashtag generation method. GT means the ground truth hashtags.}
    \label{fig:case}
\end{figure*}
Finally, we show some cases for note recommendation and generation tasks as shown in Figure~\ref{fig:case}.
In Figure~\ref{fig:case}(a), the query note suggests which summer clothes to avoid buying, while all baselines recommend summer outfits. \method can accurately recommend notes that are related to simple living.
In Figure~\ref{fig:case}(b), baselines misinterpret 'rabbit' in the note as a live rabbit, instead of the toy rabbit from \textit{Keep strangers away}.
Figure~\ref{fig:case}(c) and Figure~\ref{fig:case}(d) show the cases for hashtag generation tasks, which shows the benefits of \method.
RedHashtag is an online hashtag generation method for Xiaohongshu, which is based on classification from the fixed hashtag set.
In Figure~\ref{fig:case}(c), \method is not deceived by the semantic information 'factories'. Instead, \method correctly identifies that the note's content primarily focuses on taking photos.
In Figure~\ref{fig:case}(d), \method is capable of generating more specific and long-tail hashtags, as opposed to the more generic ones.
However, our method still suffers from the hallucination problem~\cite{yin2023woodpecker}.





\subsection{Online Experiments}
We conduct week-long online I2I recommendation experiments on Xiaohongshu.
Compared to the previous online method that adopts SentenceBERT, our NoteLLM improves the click-through rate by $16.20\%$.
Furthermore, the enhanced recall performance increases the number of comments by $1.10\%$ and the average weekly number of publishers (WAP) by $0.41\%$.
These results indicate the introduction of LLMs into I2I note recommendation tasks can improve recommendation performance and user experience.
Besides, we observe a noteworthy increase of $3.58\%$ in the number of comments on new notes within a single day.
This denotes the generalization of LLMs is beneficial to cold start notes.
Now, we have deployed our NoteLLM into the I2I note recommendation task on Xiaohongshu.

\section{Conclusion}
In this work, we propose retrievable LLMs, called \method, for note recommendation with three key components: Note Compression Prompt, GCL, and CSFT.
To manage both I2I recommendation and hashtag/category generation tasks, we utilize Note Compression Prompt to form the compressed word embeddings and generate the hashtag/category simultaneously. 
Then, we use GCL to conduct contrastive learning based on the hidden states of the compressed word, which acquires collaborative signals. 
Additionally, we employ CSFT to preserve the generation capability of \method while leveraging the semantic and collaborative information of the note to generate hashtags and categories, which can enhance the embeddings for recommendation.
Comprehensive experiments are conducted, which validate the effectiveness of \method.

\section*{Acknowledgements}
This work was supported in part by the grants from National Natural Science Foundation of China (No.62222213,
U22B2059, 62072423), and the USTC Research Funds of the Double First-Class Initiative (No.YD2150002009).

\clearpage

\bibliographystyle{ACM-Reference-Format}
\bibliography{ref}


\begin{thebibliography}{67}


\ifx \showCODEN    \undefined \def \showCODEN     #1{\unskip}     \fi
\ifx \showDOI      \undefined \def \showDOI       #1{#1}\fi
\ifx \showISBNx    \undefined \def \showISBNx     #1{\unskip}     \fi
\ifx \showISBNxiii \undefined \def \showISBNxiii  #1{\unskip}     \fi
\ifx \showISSN     \undefined \def \showISSN      #1{\unskip}     \fi
\ifx \showLCCN     \undefined \def \showLCCN      #1{\unskip}     \fi
\ifx \shownote     \undefined \def \shownote      #1{#1}          \fi
\ifx \showarticletitle \undefined \def \showarticletitle #1{#1}   \fi
\ifx \showURL      \undefined \def \showURL       {\relax}        \fi
\providecommand\bibfield[2]{#2}
\providecommand\bibinfo[2]{#2}
\providecommand\natexlab[1]{#1}
\providecommand\showeprint[2][]{arXiv:#2}

\bibitem[Bao et~al\mbox{.}(2023a)]%
        {bao2023bi}
\bibfield{author}{\bibinfo{person}{Keqin Bao}, \bibinfo{person}{Jizhi Zhang}, \bibinfo{person}{Wenjie Wang}, \bibinfo{person}{Yang Zhang}, \bibinfo{person}{Zhengyi Yang}, \bibinfo{person}{Yancheng Luo}, \bibinfo{person}{Fuli Feng}, \bibinfo{person}{Xiangnaan He}, {and} \bibinfo{person}{Qi Tian}.} \bibinfo{year}{2023}\natexlab{a}.
\newblock \showarticletitle{A bi-step grounding paradigm for large language models in recommendation systems}.
\newblock \bibinfo{journal}{\emph{arXiv preprint arXiv:2308.08434}} (\bibinfo{year}{2023}).
\newblock


\bibitem[Bao et~al\mbox{.}(2023b)]%
        {Bao_2023}
\bibfield{author}{\bibinfo{person}{Keqin Bao}, \bibinfo{person}{Jizhi Zhang}, \bibinfo{person}{Yang Zhang}, \bibinfo{person}{Wenjie Wang}, \bibinfo{person}{Fuli Feng}, {and} \bibinfo{person}{Xiangnan He}.} \bibinfo{year}{2023}\natexlab{b}.
\newblock \showarticletitle{{TALLRec}: An Effective and Efficient Tuning Framework to Align Large Language Model with Recommendation}. In \bibinfo{booktitle}{\emph{RecSys}}.
\newblock


\bibitem[Devlin et~al\mbox{.}(2018)]%
        {devlin2018bert}
\bibfield{author}{\bibinfo{person}{Jacob Devlin}, \bibinfo{person}{Ming-Wei Chang}, \bibinfo{person}{Kenton Lee}, {and} \bibinfo{person}{Kristina Toutanova}.} \bibinfo{year}{2018}\natexlab{}.
\newblock \showarticletitle{Bert: Pre-training of deep bidirectional transformers for language understanding}.
\newblock \bibinfo{journal}{\emph{arXiv preprint arXiv:1810.04805}} (\bibinfo{year}{2018}).
\newblock


\bibitem[Diao et~al\mbox{.}(2023)]%
        {diao2023hashtag}
\bibfield{author}{\bibinfo{person}{Shizhe Diao}, \bibinfo{person}{Sedrick~Scott Keh}, \bibinfo{person}{Liangming Pan}, \bibinfo{person}{Zhiliang Tian}, \bibinfo{person}{Yan Song}, {and} \bibinfo{person}{Tong Zhang}.} \bibinfo{year}{2023}\natexlab{}.
\newblock \showarticletitle{Hashtag-Guided Low-Resource Tweet Classification}. In \bibinfo{booktitle}{\emph{WWW}}. \bibinfo{pages}{1415--1426}.
\newblock


\bibitem[Fan et~al\mbox{.}(2023)]%
        {fan2023recommender}
\bibfield{author}{\bibinfo{person}{Wenqi Fan}, \bibinfo{person}{Zihuai Zhao}, \bibinfo{person}{Jiatong Li}, \bibinfo{person}{Yunqing Liu}, \bibinfo{person}{Xiaowei Mei}, \bibinfo{person}{Yiqi Wang}, \bibinfo{person}{Jiliang Tang}, {and} \bibinfo{person}{Qing Li}.} \bibinfo{year}{2023}\natexlab{}.
\newblock \showarticletitle{Recommender systems in the era of large language models (llms)}.
\newblock \bibinfo{journal}{\emph{arXiv preprint arXiv:2307.02046}} (\bibinfo{year}{2023}).
\newblock


\bibitem[He et~al\mbox{.}(2023)]%
        {he2023large}
\bibfield{author}{\bibinfo{person}{Zhankui He}, \bibinfo{person}{Zhouhang Xie}, \bibinfo{person}{Rahul Jha}, \bibinfo{person}{Harald Steck}, \bibinfo{person}{Dawen Liang}, \bibinfo{person}{Yesu Feng}, \bibinfo{person}{Bodhisattwa~Prasad Majumder}, \bibinfo{person}{Nathan Kallus}, {and} \bibinfo{person}{Julian McAuley}.} \bibinfo{year}{2023}\natexlab{}.
\newblock \showarticletitle{Large language models as zero-shot conversational recommenders}. In \bibinfo{booktitle}{\emph{CIKM}}. \bibinfo{pages}{720--730}.
\newblock


\bibitem[Hou et~al\mbox{.}(2023)]%
        {hou2023large}
\bibfield{author}{\bibinfo{person}{Yupeng Hou}, \bibinfo{person}{Junjie Zhang}, \bibinfo{person}{Zihan Lin}, \bibinfo{person}{Hongyu Lu}, \bibinfo{person}{Ruobing Xie}, \bibinfo{person}{Julian McAuley}, {and} \bibinfo{person}{Wayne~Xin Zhao}.} \bibinfo{year}{2023}\natexlab{}.
\newblock \showarticletitle{Large language models are zero-shot rankers for recommender systems}.
\newblock \bibinfo{journal}{\emph{arXiv preprint arXiv:2305.08845}} (\bibinfo{year}{2023}).
\newblock


\bibitem[Huang et~al\mbox{.}(2013)]%
        {huang2013learning}
\bibfield{author}{\bibinfo{person}{Po-Sen Huang}, \bibinfo{person}{Xiaodong He}, \bibinfo{person}{Jianfeng Gao}, \bibinfo{person}{Li Deng}, \bibinfo{person}{Alex Acero}, {and} \bibinfo{person}{Larry Heck}.} \bibinfo{year}{2013}\natexlab{}.
\newblock \showarticletitle{Learning deep structured semantic models for web search using clickthrough data}. In \bibinfo{booktitle}{\emph{CIKM}}. \bibinfo{pages}{2333--2338}.
\newblock


\bibitem[Huang et~al\mbox{.}(2023)]%
        {huang2023recommender}
\bibfield{author}{\bibinfo{person}{Xu Huang}, \bibinfo{person}{Jianxun Lian}, \bibinfo{person}{Yuxuan Lei}, \bibinfo{person}{Jing Yao}, \bibinfo{person}{Defu Lian}, {and} \bibinfo{person}{Xing Xie}.} \bibinfo{year}{2023}\natexlab{}.
\newblock \showarticletitle{Recommender AI Agent: Integrating Large Language Models for Interactive Recommendations}.
\newblock \bibinfo{journal}{\emph{arXiv preprint arXiv:2308.16505}} (\bibinfo{year}{2023}).
\newblock


\bibitem[Jiang et~al\mbox{.}(2023)]%
        {jiang2023scaling}
\bibfield{author}{\bibinfo{person}{Ting Jiang}, \bibinfo{person}{Shaohan Huang}, \bibinfo{person}{Zhongzhi Luan}, \bibinfo{person}{Deqing Wang}, {and} \bibinfo{person}{Fuzhen Zhuang}.} \bibinfo{year}{2023}\natexlab{}.
\newblock \showarticletitle{Scaling Sentence Embeddings with Large Language Models}.
\newblock \bibinfo{journal}{\emph{arXiv preprint arXiv:2307.16645}} (\bibinfo{year}{2023}).
\newblock


\bibitem[Jiang et~al\mbox{.}(2022)]%
        {jiang2022promptbert}
\bibfield{author}{\bibinfo{person}{Ting Jiang}, \bibinfo{person}{Jian Jiao}, \bibinfo{person}{Shaohan Huang}, \bibinfo{person}{Zihan Zhang}, \bibinfo{person}{Deqing Wang}, \bibinfo{person}{Fuzhen Zhuang}, \bibinfo{person}{Furu Wei}, \bibinfo{person}{Haizhen Huang}, \bibinfo{person}{Denvy Deng}, {and} \bibinfo{person}{Qi Zhang}.} \bibinfo{year}{2022}\natexlab{}.
\newblock \showarticletitle{PromptBERT: Improving BERT Sentence Embeddings with Prompts}. In \bibinfo{booktitle}{\emph{EMNLP}}. \bibinfo{pages}{8826--8837}.
\newblock


\bibitem[Johnson et~al\mbox{.}(2019)]%
        {johnson2019billion}
\bibfield{author}{\bibinfo{person}{Jeff Johnson}, \bibinfo{person}{Matthijs Douze}, {and} \bibinfo{person}{Herv{\'e} J{\'e}gou}.} \bibinfo{year}{2019}\natexlab{}.
\newblock \showarticletitle{Billion-scale similarity search with {GPUs}}.
\newblock \bibinfo{journal}{\emph{IEEE Transactions on Big Data}} \bibinfo{volume}{7}, \bibinfo{number}{3} (\bibinfo{year}{2019}), \bibinfo{pages}{535--547}.
\newblock


\bibitem[Karpukhin et~al\mbox{.}(2020)]%
        {karpukhin2020dense}
\bibfield{author}{\bibinfo{person}{Vladimir Karpukhin}, \bibinfo{person}{Barlas O{\u{g}}uz}, \bibinfo{person}{Sewon Min}, \bibinfo{person}{Patrick Lewis}, \bibinfo{person}{Ledell Wu}, \bibinfo{person}{Sergey Edunov}, \bibinfo{person}{Danqi Chen}, {and} \bibinfo{person}{Wen-tau Yih}.} \bibinfo{year}{2020}\natexlab{}.
\newblock \showarticletitle{Dense passage retrieval for open-domain question answering}.
\newblock \bibinfo{journal}{\emph{arXiv preprint arXiv:2004.04906}} (\bibinfo{year}{2020}).
\newblock


\bibitem[Kou et~al\mbox{.}(2018)]%
        {kou2018hashtag}
\bibfield{author}{\bibinfo{person}{Fei-Fei Kou}, \bibinfo{person}{Jun-Ping Du}, \bibinfo{person}{Cong-Xian Yang}, \bibinfo{person}{Yan-Song Shi}, \bibinfo{person}{Wan-Qiu Cui}, \bibinfo{person}{Mei-Yu Liang}, {and} \bibinfo{person}{Yue Geng}.} \bibinfo{year}{2018}\natexlab{}.
\newblock \showarticletitle{Hashtag recommendation based on multi-features of microblogs}.
\newblock \bibinfo{journal}{\emph{JCST}}  \bibinfo{volume}{33} (\bibinfo{year}{2018}), \bibinfo{pages}{711--726}.
\newblock


\bibitem[Li et~al\mbox{.}(2023)]%
        {li2023exploring}
\bibfield{author}{\bibinfo{person}{Ruyu Li}, \bibinfo{person}{Wenhao Deng}, \bibinfo{person}{Yu Cheng}, \bibinfo{person}{Zheng Yuan}, \bibinfo{person}{Jiaqi Zhang}, {and} \bibinfo{person}{Fajie Yuan}.} \bibinfo{year}{2023}\natexlab{}.
\newblock \showarticletitle{Exploring the Upper Limits of Text-Based Collaborative Filtering Using Large Language Models: Discoveries and Insights}.
\newblock \bibinfo{journal}{\emph{arXiv preprint arXiv:2305.11700}} (\bibinfo{year}{2023}).
\newblock


\bibitem[Liao et~al\mbox{.}(2022)]%
        {liao2022user}
\bibfield{author}{\bibinfo{person}{Mengqi Liao}, \bibinfo{person}{S~Shyam Sundar}, {and} \bibinfo{person}{Joseph B.~Walther}.} \bibinfo{year}{2022}\natexlab{}.
\newblock \showarticletitle{User trust in recommendation systems: A comparison of content-based, collaborative and demographic filtering}. In \bibinfo{booktitle}{\emph{CHI}}. \bibinfo{pages}{1--14}.
\newblock


\bibitem[Lin et~al\mbox{.}(2022)]%
        {lin2022quantifying}
\bibfield{author}{\bibinfo{person}{Allen Lin}, \bibinfo{person}{Jianling Wang}, \bibinfo{person}{Ziwei Zhu}, {and} \bibinfo{person}{James Caverlee}.} \bibinfo{year}{2022}\natexlab{}.
\newblock \showarticletitle{Quantifying and mitigating popularity bias in conversational recommender systems}. In \bibinfo{booktitle}{\emph{CIKM}}. \bibinfo{pages}{1238--1247}.
\newblock


\bibitem[Lin et~al\mbox{.}(2023)]%
        {lin2023can}
\bibfield{author}{\bibinfo{person}{Jianghao Lin}, \bibinfo{person}{Xinyi Dai}, \bibinfo{person}{Yunjia Xi}, \bibinfo{person}{Weiwen Liu}, \bibinfo{person}{Bo Chen}, \bibinfo{person}{Xiangyang Li}, \bibinfo{person}{Chenxu Zhu}, \bibinfo{person}{Huifeng Guo}, \bibinfo{person}{Yong Yu}, \bibinfo{person}{Ruiming Tang}, {et~al\mbox{.}}} \bibinfo{year}{2023}\natexlab{}.
\newblock \showarticletitle{How Can Recommender Systems Benefit from Large Language Models: A Survey}.
\newblock \bibinfo{journal}{\emph{arXiv preprint arXiv:2306.05817}} (\bibinfo{year}{2023}).
\newblock


\bibitem[Linden et~al\mbox{.}(2003)]%
        {linden2003amazon}
\bibfield{author}{\bibinfo{person}{Greg Linden}, \bibinfo{person}{Brent Smith}, {and} \bibinfo{person}{Jeremy York}.} \bibinfo{year}{2003}\natexlab{}.
\newblock \showarticletitle{Amazon. com recommendations: Item-to-item collaborative filtering}.
\newblock \bibinfo{journal}{\emph{IEEE Internet computing}} \bibinfo{volume}{7}, \bibinfo{number}{1} (\bibinfo{year}{2003}), \bibinfo{pages}{76--80}.
\newblock


\bibitem[Liu et~al\mbox{.}(2023b)]%
        {liu2023chatgpt}
\bibfield{author}{\bibinfo{person}{Junling Liu}, \bibinfo{person}{Chao Liu}, \bibinfo{person}{Renjie Lv}, \bibinfo{person}{Kang Zhou}, {and} \bibinfo{person}{Yan Zhang}.} \bibinfo{year}{2023}\natexlab{b}.
\newblock \showarticletitle{Is chatgpt a good recommender? a preliminary study}.
\newblock \bibinfo{journal}{\emph{arXiv preprint arXiv:2304.10149}} (\bibinfo{year}{2023}).
\newblock


\bibitem[Liu et~al\mbox{.}(2023a)]%
        {liu2023first}
\bibfield{author}{\bibinfo{person}{Qijiong Liu}, \bibinfo{person}{Nuo Chen}, \bibinfo{person}{Tetsuya Sakai}, {and} \bibinfo{person}{Xiao-Ming Wu}.} \bibinfo{year}{2023}\natexlab{a}.
\newblock \showarticletitle{A First Look at LLM-Powered Generative News Recommendation}.
\newblock \bibinfo{journal}{\emph{arXiv preprint arXiv:2305.06566}} (\bibinfo{year}{2023}).
\newblock


\bibitem[Longpre et~al\mbox{.}(2023)]%
        {longpre2023flan}
\bibfield{author}{\bibinfo{person}{Shayne Longpre}, \bibinfo{person}{Le Hou}, \bibinfo{person}{Tu Vu}, \bibinfo{person}{Albert Webson}, \bibinfo{person}{Hyung~Won Chung}, \bibinfo{person}{Yi Tay}, \bibinfo{person}{Denny Zhou}, \bibinfo{person}{Quoc~V Le}, \bibinfo{person}{Barret Zoph}, \bibinfo{person}{Jason Wei}, {et~al\mbox{.}}} \bibinfo{year}{2023}\natexlab{}.
\newblock \showarticletitle{The flan collection: Designing data and methods for effective instruction tuning}.
\newblock \bibinfo{journal}{\emph{arXiv preprint arXiv:2301.13688}} (\bibinfo{year}{2023}).
\newblock


\bibitem[Lyu et~al\mbox{.}(2023)]%
        {lyu2023llm}
\bibfield{author}{\bibinfo{person}{Hanjia Lyu}, \bibinfo{person}{Song Jiang}, \bibinfo{person}{Hanqing Zeng}, \bibinfo{person}{Yinglong Xia}, {and} \bibinfo{person}{Jiebo Luo}.} \bibinfo{year}{2023}\natexlab{}.
\newblock \showarticletitle{Llm-rec: Personalized recommendation via prompting large language models}.
\newblock \bibinfo{journal}{\emph{arXiv preprint arXiv:2307.15780}} (\bibinfo{year}{2023}).
\newblock


\bibitem[Lyu et~al\mbox{.}(2024)]%
        {lyu2024crud}
\bibfield{author}{\bibinfo{person}{Yuanjie Lyu}, \bibinfo{person}{Zhiyu Li}, \bibinfo{person}{Simin Niu}, \bibinfo{person}{Feiyu Xiong}, \bibinfo{person}{Bo Tang}, \bibinfo{person}{Wenjin Wang}, \bibinfo{person}{Hao Wu}, \bibinfo{person}{Huanyong Liu}, \bibinfo{person}{Tong Xu}, {and} \bibinfo{person}{Enhong Chen}.} \bibinfo{year}{2024}\natexlab{}.
\newblock \showarticletitle{CRUD-RAG: A Comprehensive Chinese Benchmark for Retrieval-Augmented Generation of Large Language Models}.
\newblock \bibinfo{journal}{\emph{arXiv preprint arXiv:2401.17043}} (\bibinfo{year}{2024}).
\newblock


\bibitem[Ma et~al\mbox{.}(2022)]%
        {ma2022pre}
\bibfield{author}{\bibinfo{person}{Xinyu Ma}, \bibinfo{person}{Jiafeng Guo}, \bibinfo{person}{Ruqing Zhang}, \bibinfo{person}{Yixing Fan}, {and} \bibinfo{person}{Xueqi Cheng}.} \bibinfo{year}{2022}\natexlab{}.
\newblock \showarticletitle{Pre-train a discriminative text encoder for dense retrieval via contrastive span prediction}. In \bibinfo{booktitle}{\emph{SIGIR}}. \bibinfo{pages}{848--858}.
\newblock


\bibitem[Ma et~al\mbox{.}(2023)]%
        {ma2023fine}
\bibfield{author}{\bibinfo{person}{Xueguang Ma}, \bibinfo{person}{Liang Wang}, \bibinfo{person}{Nan Yang}, \bibinfo{person}{Furu Wei}, {and} \bibinfo{person}{Jimmy Lin}.} \bibinfo{year}{2023}\natexlab{}.
\newblock \showarticletitle{Fine-Tuning LLaMA for Multi-Stage Text Retrieval}.
\newblock \bibinfo{journal}{\emph{arXiv preprint arXiv:2310.08319}} (\bibinfo{year}{2023}).
\newblock


\bibitem[Mikolov et~al\mbox{.}(2013)]%
        {mikolov2013efficient}
\bibfield{author}{\bibinfo{person}{Tomas Mikolov}, \bibinfo{person}{Kai Chen}, \bibinfo{person}{Greg Corrado}, {and} \bibinfo{person}{Jeffrey Dean}.} \bibinfo{year}{2013}\natexlab{}.
\newblock \showarticletitle{Efficient estimation of word representations in vector space}.
\newblock \bibinfo{journal}{\emph{arXiv preprint arXiv:1301.3781}} (\bibinfo{year}{2013}).
\newblock


\bibitem[Muennighoff(2022)]%
        {muennighoff2022sgpt}
\bibfield{author}{\bibinfo{person}{Niklas Muennighoff}.} \bibinfo{year}{2022}\natexlab{}.
\newblock \showarticletitle{Sgpt: Gpt sentence embeddings for semantic search}.
\newblock \bibinfo{journal}{\emph{arXiv preprint arXiv:2202.08904}} (\bibinfo{year}{2022}).
\newblock


\bibitem[Mysore et~al\mbox{.}(2023)]%
        {mysore2023large}
\bibfield{author}{\bibinfo{person}{Sheshera Mysore}, \bibinfo{person}{Andrew McCallum}, {and} \bibinfo{person}{Hamed Zamani}.} \bibinfo{year}{2023}\natexlab{}.
\newblock \showarticletitle{Large Language Model Augmented Narrative Driven Recommendations}.
\newblock \bibinfo{journal}{\emph{arXiv preprint arXiv:2306.02250}} (\bibinfo{year}{2023}).
\newblock


\bibitem[Neelakantan et~al\mbox{.}(2022)]%
        {neelakantan2022text}
\bibfield{author}{\bibinfo{person}{Arvind Neelakantan}, \bibinfo{person}{Tao Xu}, \bibinfo{person}{Raul Puri}, \bibinfo{person}{Alec Radford}, \bibinfo{person}{Jesse~Michael Han}, \bibinfo{person}{Jerry Tworek}, \bibinfo{person}{Qiming Yuan}, \bibinfo{person}{Nikolas Tezak}, \bibinfo{person}{Jong~Wook Kim}, \bibinfo{person}{Chris Hallacy}, {et~al\mbox{.}}} \bibinfo{year}{2022}\natexlab{}.
\newblock \showarticletitle{Text and code embeddings by contrastive pre-training}.
\newblock \bibinfo{journal}{\emph{arXiv preprint arXiv:2201.10005}} (\bibinfo{year}{2022}).
\newblock


\bibitem[OpenAI(2023)]%
        {openai2023gpt4}
\bibfield{author}{\bibinfo{person}{OpenAI}.} \bibinfo{year}{2023}\natexlab{}.
\newblock \showarticletitle{GPT-4 Technical Report}.
\newblock \bibinfo{journal}{\emph{arXiv preprint arXiv:2303.08774}} (\bibinfo{year}{2023}).
\newblock


\bibitem[Ouyang et~al\mbox{.}(2022)]%
        {ouyang2022training}
\bibfield{author}{\bibinfo{person}{Long Ouyang}, \bibinfo{person}{Jeffrey Wu}, \bibinfo{person}{Xu Jiang}, \bibinfo{person}{Diogo Almeida}, \bibinfo{person}{Carroll Wainwright}, \bibinfo{person}{Pamela Mishkin}, \bibinfo{person}{Chong Zhang}, \bibinfo{person}{Sandhini Agarwal}, \bibinfo{person}{Katarina Slama}, \bibinfo{person}{Alex Ray}, {et~al\mbox{.}}} \bibinfo{year}{2022}\natexlab{}.
\newblock \showarticletitle{Training language models to follow instructions with human feedback}.
\newblock \bibinfo{journal}{\emph{NeurIPS}}  \bibinfo{volume}{35} (\bibinfo{year}{2022}), \bibinfo{pages}{27730--27744}.
\newblock


\bibitem[Peng et~al\mbox{.}(2023a)]%
        {peng2023large}
\bibfield{author}{\bibinfo{person}{Wenjun Peng}, \bibinfo{person}{Guiyang Li}, \bibinfo{person}{Yue Jiang}, \bibinfo{person}{Zilong Wang}, \bibinfo{person}{Dan Ou}, \bibinfo{person}{Xiaoyi Zeng}, \bibinfo{person}{Enhong Chen}, {et~al\mbox{.}}} \bibinfo{year}{2023}\natexlab{a}.
\newblock \showarticletitle{Large Language Model based Long-tail Query Rewriting in Taobao Search}.
\newblock \bibinfo{journal}{\emph{arXiv preprint arXiv:2311.03758}} (\bibinfo{year}{2023}).
\newblock


\bibitem[Peng et~al\mbox{.}(2023b)]%
        {peng2023gpt}
\bibfield{author}{\bibinfo{person}{Wenjun Peng}, \bibinfo{person}{Derong Xu}, \bibinfo{person}{Tong Xu}, \bibinfo{person}{Jianjin Zhang}, {and} \bibinfo{person}{Enhong Chen}.} \bibinfo{year}{2023}\natexlab{b}.
\newblock \showarticletitle{Are gpt embeddings useful for ads and recommendation?}. In \bibinfo{booktitle}{\emph{KSEM}}. Springer, \bibinfo{pages}{151--162}.
\newblock


\bibitem[Ramos et~al\mbox{.}(2003)]%
        {ramos2003using}
\bibfield{author}{\bibinfo{person}{Juan Ramos} {et~al\mbox{.}}} \bibinfo{year}{2003}\natexlab{}.
\newblock \showarticletitle{Using tf-idf to determine word relevance in document queries}. In \bibinfo{booktitle}{\emph{Proceedings of the first instructional conference on machine learning}}, Vol.~\bibinfo{volume}{242}. Citeseer, \bibinfo{pages}{29--48}.
\newblock


\bibitem[Reimers and Gurevych(2019)]%
        {reimers2019sentence}
\bibfield{author}{\bibinfo{person}{Nils Reimers} {and} \bibinfo{person}{Iryna Gurevych}.} \bibinfo{year}{2019}\natexlab{}.
\newblock \showarticletitle{Sentence-BERT: Sentence Embeddings using Siamese BERT-Networks}. In \bibinfo{booktitle}{\emph{EMNLP-IJCNLP}}.
\newblock


\bibitem[Robertson et~al\mbox{.}(2009)]%
        {robertson2009probabilistic}
\bibfield{author}{\bibinfo{person}{Stephen Robertson}, \bibinfo{person}{Hugo Zaragoza}, {et~al\mbox{.}}} \bibinfo{year}{2009}\natexlab{}.
\newblock \showarticletitle{The probabilistic relevance framework: BM25 and beyond}.
\newblock \bibinfo{journal}{\emph{Foundations and Trends{\textregistered} in Information Retrieval}} \bibinfo{volume}{3}, \bibinfo{number}{4} (\bibinfo{year}{2009}), \bibinfo{pages}{333--389}.
\newblock


\bibitem[Stiennon et~al\mbox{.}(2020)]%
        {stiennon2020learning}
\bibfield{author}{\bibinfo{person}{Nisan Stiennon}, \bibinfo{person}{Long Ouyang}, \bibinfo{person}{Jeffrey Wu}, \bibinfo{person}{Daniel Ziegler}, \bibinfo{person}{Ryan Lowe}, \bibinfo{person}{Chelsea Voss}, \bibinfo{person}{Alec Radford}, \bibinfo{person}{Dario Amodei}, {and} \bibinfo{person}{Paul~F Christiano}.} \bibinfo{year}{2020}\natexlab{}.
\newblock \showarticletitle{Learning to summarize with human feedback}.
\newblock \bibinfo{journal}{\emph{NeurIPS}}  \bibinfo{volume}{33} (\bibinfo{year}{2020}), \bibinfo{pages}{3008--3021}.
\newblock


\bibitem[Su et~al\mbox{.}(2023)]%
        {su2022one}
\bibfield{author}{\bibinfo{person}{Hongjin Su}, \bibinfo{person}{Jungo Kasai}, \bibinfo{person}{Yizhong Wang}, \bibinfo{person}{Yushi Hu}, \bibinfo{person}{Mari Ostendorf}, \bibinfo{person}{Wen-tau Yih}, \bibinfo{person}{Noah~A Smith}, \bibinfo{person}{Luke Zettlemoyer}, \bibinfo{person}{Tao Yu}, {et~al\mbox{.}}} \bibinfo{year}{2023}\natexlab{}.
\newblock \showarticletitle{One embedder, any task: Instruction-finetuned text embeddings}.
\newblock \bibinfo{journal}{\emph{ACL Findings}} (\bibinfo{year}{2023}).
\newblock


\bibitem[Toneva et~al\mbox{.}(2018)]%
        {toneva2018empirical}
\bibfield{author}{\bibinfo{person}{Mariya Toneva}, \bibinfo{person}{Alessandro Sordoni}, \bibinfo{person}{Remi Tachet~des Combes}, \bibinfo{person}{Adam Trischler}, \bibinfo{person}{Yoshua Bengio}, {and} \bibinfo{person}{Geoffrey~J Gordon}.} \bibinfo{year}{2018}\natexlab{}.
\newblock \showarticletitle{An empirical study of example forgetting during deep neural network learning}.
\newblock \bibinfo{journal}{\emph{arXiv preprint arXiv:1812.05159}} (\bibinfo{year}{2018}).
\newblock


\bibitem[Touvron et~al\mbox{.}(2023a)]%
        {touvron2023llama}
\bibfield{author}{\bibinfo{person}{Hugo Touvron}, \bibinfo{person}{Thibaut Lavril}, \bibinfo{person}{Gautier Izacard}, \bibinfo{person}{Xavier Martinet}, \bibinfo{person}{Marie-Anne Lachaux}, \bibinfo{person}{Timoth{\'e}e Lacroix}, \bibinfo{person}{Baptiste Rozi{\`e}re}, \bibinfo{person}{Naman Goyal}, \bibinfo{person}{Eric Hambro}, \bibinfo{person}{Faisal Azhar}, {et~al\mbox{.}}} \bibinfo{year}{2023}\natexlab{a}.
\newblock \showarticletitle{Llama: Open and efficient foundation language models}.
\newblock \bibinfo{journal}{\emph{arXiv preprint arXiv:2302.13971}} (\bibinfo{year}{2023}).
\newblock


\bibitem[Touvron et~al\mbox{.}(2023b)]%
        {touvron2023llama2}
\bibfield{author}{\bibinfo{person}{Hugo Touvron}, \bibinfo{person}{Louis Martin}, \bibinfo{person}{Kevin Stone}, \bibinfo{person}{Peter Albert}, \bibinfo{person}{Amjad Almahairi}, \bibinfo{person}{Yasmine Babaei}, \bibinfo{person}{Nikolay Bashlykov}, \bibinfo{person}{Soumya Batra}, \bibinfo{person}{Prajjwal Bhargava}, \bibinfo{person}{Shruti Bhosale}, {et~al\mbox{.}}} \bibinfo{year}{2023}\natexlab{b}.
\newblock \showarticletitle{Llama 2: Open foundation and fine-tuned chat models}.
\newblock \bibinfo{journal}{\emph{arXiv preprint arXiv:2307.09288}} (\bibinfo{year}{2023}).
\newblock


\bibitem[Wang et~al\mbox{.}(2023b)]%
        {wang2023recmind}
\bibfield{author}{\bibinfo{person}{Yancheng Wang}, \bibinfo{person}{Ziyan Jiang}, \bibinfo{person}{Zheng Chen}, \bibinfo{person}{Fan Yang}, \bibinfo{person}{Yingxue Zhou}, \bibinfo{person}{Eunah Cho}, \bibinfo{person}{Xing Fan}, \bibinfo{person}{Xiaojiang Huang}, \bibinfo{person}{Yanbin Lu}, {and} \bibinfo{person}{Yingzhen Yang}.} \bibinfo{year}{2023}\natexlab{b}.
\newblock \showarticletitle{Recmind: Large language model powered agent for recommendation}.
\newblock \bibinfo{journal}{\emph{arXiv preprint arXiv:2308.14296}} (\bibinfo{year}{2023}).
\newblock


\bibitem[Wang et~al\mbox{.}(2019a)]%
        {wang2019topic}
\bibfield{author}{\bibinfo{person}{Yue Wang}, \bibinfo{person}{Jing Li}, \bibinfo{person}{Hou~Pong Chan}, \bibinfo{person}{Irwin King}, \bibinfo{person}{Michael~R Lyu}, {and} \bibinfo{person}{Shuming Shi}.} \bibinfo{year}{2019}\natexlab{a}.
\newblock \showarticletitle{Topic-Aware Neural Keyphrase Generation for Social Media Language}. In \bibinfo{booktitle}{\emph{ACL}}. \bibinfo{pages}{2516--2526}.
\newblock


\bibitem[Wang et~al\mbox{.}(2019b)]%
        {wang2019microblog}
\bibfield{author}{\bibinfo{person}{Yue Wang}, \bibinfo{person}{Jing Li}, \bibinfo{person}{Irwin King}, \bibinfo{person}{Michael~R Lyu}, {and} \bibinfo{person}{Shuming Shi}.} \bibinfo{year}{2019}\natexlab{b}.
\newblock \showarticletitle{Microblog hashtag generation via encoding conversation contexts}.
\newblock \bibinfo{journal}{\emph{arXiv preprint arXiv:1905.07584}} (\bibinfo{year}{2019}).
\newblock


\bibitem[Wang et~al\mbox{.}(2023a)]%
        {wang2023anypredict}
\bibfield{author}{\bibinfo{person}{Zifeng Wang}, \bibinfo{person}{Chufan Gao}, \bibinfo{person}{Cao Xiao}, {and} \bibinfo{person}{Jimeng Sun}.} \bibinfo{year}{2023}\natexlab{a}.
\newblock \showarticletitle{AnyPredict: Foundation Model for Tabular Prediction}.
\newblock \bibinfo{journal}{\emph{arXiv preprint arXiv:2305.12081}} (\bibinfo{year}{2023}).
\newblock


\bibitem[Wei et~al\mbox{.}(2021)]%
        {wei2021finetuned}
\bibfield{author}{\bibinfo{person}{Jason Wei}, \bibinfo{person}{Maarten Bosma}, \bibinfo{person}{Vincent~Y Zhao}, \bibinfo{person}{Kelvin Guu}, \bibinfo{person}{Adams~Wei Yu}, \bibinfo{person}{Brian Lester}, \bibinfo{person}{Nan Du}, \bibinfo{person}{Andrew~M Dai}, {and} \bibinfo{person}{Quoc~V Le}.} \bibinfo{year}{2021}\natexlab{}.
\newblock \showarticletitle{Finetuned language models are zero-shot learners}.
\newblock \bibinfo{journal}{\emph{arXiv preprint arXiv:2109.01652}} (\bibinfo{year}{2021}).
\newblock


\bibitem[Wu et~al\mbox{.}(2023a)]%
        {wu2023personalized}
\bibfield{author}{\bibinfo{person}{Chuhan Wu}, \bibinfo{person}{Fangzhao Wu}, \bibinfo{person}{Yongfeng Huang}, {and} \bibinfo{person}{Xing Xie}.} \bibinfo{year}{2023}\natexlab{a}.
\newblock \showarticletitle{Personalized news recommendation: Methods and challenges}.
\newblock \bibinfo{journal}{\emph{ACM Transactions on Information Systems}} \bibinfo{volume}{41}, \bibinfo{number}{1} (\bibinfo{year}{2023}), \bibinfo{pages}{1--50}.
\newblock


\bibitem[Wu et~al\mbox{.}(2021)]%
        {wu2021empowering}
\bibfield{author}{\bibinfo{person}{Chuhan Wu}, \bibinfo{person}{Fangzhao Wu}, \bibinfo{person}{Tao Qi}, {and} \bibinfo{person}{Yongfeng Huang}.} \bibinfo{year}{2021}\natexlab{}.
\newblock \showarticletitle{Empowering news recommendation with pre-trained language models}. In \bibinfo{booktitle}{\emph{SIGIR}}. \bibinfo{pages}{1652--1656}.
\newblock


\bibitem[Wu et~al\mbox{.}(2023b)]%
        {wu2023survey}
\bibfield{author}{\bibinfo{person}{Likang Wu}, \bibinfo{person}{Zhi Zheng}, \bibinfo{person}{Zhaopeng Qiu}, \bibinfo{person}{Hao Wang}, \bibinfo{person}{Hongchao Gu}, \bibinfo{person}{Tingjia Shen}, \bibinfo{person}{Chuan Qin}, \bibinfo{person}{Chen Zhu}, \bibinfo{person}{Hengshu Zhu}, \bibinfo{person}{Qi Liu}, {et~al\mbox{.}}} \bibinfo{year}{2023}\natexlab{b}.
\newblock \showarticletitle{A Survey on Large Language Models for Recommendation}.
\newblock \bibinfo{journal}{\emph{arXiv preprint arXiv:2305.19860}} (\bibinfo{year}{2023}).
\newblock


\bibitem[Xi et~al\mbox{.}(2023)]%
        {xi2023towards}
\bibfield{author}{\bibinfo{person}{Yunjia Xi}, \bibinfo{person}{Weiwen Liu}, \bibinfo{person}{Jianghao Lin}, \bibinfo{person}{Jieming Zhu}, \bibinfo{person}{Bo Chen}, \bibinfo{person}{Ruiming Tang}, \bibinfo{person}{Weinan Zhang}, \bibinfo{person}{Rui Zhang}, {and} \bibinfo{person}{Yong Yu}.} \bibinfo{year}{2023}\natexlab{}.
\newblock \showarticletitle{Towards Open-World Recommendation with Knowledge Augmentation from Large Language Models}.
\newblock \bibinfo{journal}{\emph{arXiv preprint arXiv:2306.10933}} (\bibinfo{year}{2023}).
\newblock


\bibitem[Xiao et~al\mbox{.}(2022a)]%
        {xiao2022progressively}
\bibfield{author}{\bibinfo{person}{Shitao Xiao}, \bibinfo{person}{Zheng Liu}, \bibinfo{person}{Weihao Han}, \bibinfo{person}{Jianjin Zhang}, \bibinfo{person}{Yingxia Shao}, \bibinfo{person}{Defu Lian}, \bibinfo{person}{Chaozhuo Li}, \bibinfo{person}{Hao Sun}, \bibinfo{person}{Denvy Deng}, \bibinfo{person}{Liangjie Zhang}, {et~al\mbox{.}}} \bibinfo{year}{2022}\natexlab{a}.
\newblock \showarticletitle{Progressively optimized bi-granular document representation for scalable embedding based retrieval}. In \bibinfo{booktitle}{\emph{WWW}}. \bibinfo{pages}{286--296}.
\newblock


\bibitem[Xiao et~al\mbox{.}(2022b)]%
        {xiao2022training}
\bibfield{author}{\bibinfo{person}{Shitao Xiao}, \bibinfo{person}{Zheng Liu}, \bibinfo{person}{Yingxia Shao}, \bibinfo{person}{Tao Di}, \bibinfo{person}{Bhuvan Middha}, \bibinfo{person}{Fangzhao Wu}, {and} \bibinfo{person}{Xing Xie}.} \bibinfo{year}{2022}\natexlab{b}.
\newblock \showarticletitle{Training large-scale news recommenders with pretrained language models in the loop}. In \bibinfo{booktitle}{\emph{KDD}}. \bibinfo{pages}{4215--4225}.
\newblock


\bibitem[Xu et~al\mbox{.}(2023)]%
        {xu2023large}
\bibfield{author}{\bibinfo{person}{Derong Xu}, \bibinfo{person}{Wei Chen}, \bibinfo{person}{Wenjun Peng}, \bibinfo{person}{Chao Zhang}, \bibinfo{person}{Tong Xu}, \bibinfo{person}{Xiangyu Zhao}, \bibinfo{person}{Xian Wu}, \bibinfo{person}{Yefeng Zheng}, {and} \bibinfo{person}{Enhong Chen}.} \bibinfo{year}{2023}\natexlab{}.
\newblock \showarticletitle{Large Language Models for Generative Information Extraction: A Survey}.
\newblock \bibinfo{journal}{\emph{arXiv preprint arXiv:2312.17617}} (\bibinfo{year}{2023}).
\newblock


\bibitem[Yang et~al\mbox{.}(2020)]%
        {yang2020large}
\bibfield{author}{\bibinfo{person}{Xiaoyong Yang}, \bibinfo{person}{Yadong Zhu}, \bibinfo{person}{Yi Zhang}, \bibinfo{person}{Xiaobo Wang}, {and} \bibinfo{person}{Quan Yuan}.} \bibinfo{year}{2020}\natexlab{}.
\newblock \showarticletitle{Large scale product graph construction for recommendation in e-commerce}.
\newblock \bibinfo{journal}{\emph{arXiv preprint arXiv:2010.05525}} (\bibinfo{year}{2020}).
\newblock


\bibitem[Yin et~al\mbox{.}(2023a)]%
        {yin2023survey}
\bibfield{author}{\bibinfo{person}{Shukang Yin}, \bibinfo{person}{Chaoyou Fu}, \bibinfo{person}{Sirui Zhao}, \bibinfo{person}{Ke Li}, \bibinfo{person}{Xing Sun}, \bibinfo{person}{Tong Xu}, {and} \bibinfo{person}{Enhong Chen}.} \bibinfo{year}{2023}\natexlab{a}.
\newblock \showarticletitle{A Survey on Multimodal Large Language Models}.
\newblock \bibinfo{journal}{\emph{arXiv preprint arXiv:2306.13549}} (\bibinfo{year}{2023}).
\newblock


\bibitem[Yin et~al\mbox{.}(2023b)]%
        {yin2023woodpecker}
\bibfield{author}{\bibinfo{person}{Shukang Yin}, \bibinfo{person}{Chaoyou Fu}, \bibinfo{person}{Sirui Zhao}, \bibinfo{person}{Tong Xu}, \bibinfo{person}{Hao Wang}, \bibinfo{person}{Dianbo Sui}, \bibinfo{person}{Yunhang Shen}, \bibinfo{person}{Ke Li}, \bibinfo{person}{Xing Sun}, {and} \bibinfo{person}{Enhong Chen}.} \bibinfo{year}{2023}\natexlab{b}.
\newblock \showarticletitle{Woodpecker: Hallucination correction for multimodal large language models}.
\newblock \bibinfo{journal}{\emph{arXiv preprint arXiv:2310.16045}} (\bibinfo{year}{2023}).
\newblock


\bibitem[Zeng et~al\mbox{.}(2018)]%
        {zeng2018topic}
\bibfield{author}{\bibinfo{person}{Jichuan Zeng}, \bibinfo{person}{Jing Li}, \bibinfo{person}{Yan Song}, \bibinfo{person}{Cuiyun Gao}, \bibinfo{person}{Michael~R Lyu}, {and} \bibinfo{person}{Irwin King}.} \bibinfo{year}{2018}\natexlab{}.
\newblock \showarticletitle{Topic Memory Networks for Short Text Classification}. In \bibinfo{booktitle}{\emph{EMNLP}}. \bibinfo{pages}{3120--3131}.
\newblock


\bibitem[Zhang et~al\mbox{.}(2023b)]%
        {zhang2023recommendation}
\bibfield{author}{\bibinfo{person}{Junjie Zhang}, \bibinfo{person}{Ruobing Xie}, \bibinfo{person}{Yupeng Hou}, \bibinfo{person}{Wayne~Xin Zhao}, \bibinfo{person}{Leyu Lin}, {and} \bibinfo{person}{Ji-Rong Wen}.} \bibinfo{year}{2023}\natexlab{b}.
\newblock \showarticletitle{Recommendation as instruction following: A large language model empowered recommendation approach}.
\newblock \bibinfo{journal}{\emph{arXiv preprint arXiv:2305.07001}} (\bibinfo{year}{2023}).
\newblock


\bibitem[Zhang et~al\mbox{.}(2017)]%
        {zhang2017hashtag}
\bibfield{author}{\bibinfo{person}{Qi Zhang}, \bibinfo{person}{Jiawen Wang}, \bibinfo{person}{Haoran Huang}, \bibinfo{person}{Xuanjing Huang}, {and} \bibinfo{person}{Yeyun Gong}.} \bibinfo{year}{2017}\natexlab{}.
\newblock \showarticletitle{Hashtag Recommendation for Multimodal Microblog Using Co-Attention Network.}. In \bibinfo{booktitle}{\emph{IJCAI}}. \bibinfo{pages}{3420--3426}.
\newblock


\bibitem[Zhang et~al\mbox{.}(2016)]%
        {zhang2016keyphrase}
\bibfield{author}{\bibinfo{person}{Qi Zhang}, \bibinfo{person}{Yang Wang}, \bibinfo{person}{Yeyun Gong}, {and} \bibinfo{person}{Xuan-Jing Huang}.} \bibinfo{year}{2016}\natexlab{}.
\newblock \showarticletitle{Keyphrase extraction using deep recurrent neural networks on twitter}. In \bibinfo{booktitle}{\emph{EMNLP}}. \bibinfo{pages}{836--845}.
\newblock


\bibitem[Zhang et~al\mbox{.}(2023a)]%
        {zhang2023instruction}
\bibfield{author}{\bibinfo{person}{Shengyu Zhang}, \bibinfo{person}{Linfeng Dong}, \bibinfo{person}{Xiaoya Li}, \bibinfo{person}{Sen Zhang}, \bibinfo{person}{Xiaofei Sun}, \bibinfo{person}{Shuhe Wang}, \bibinfo{person}{Jiwei Li}, \bibinfo{person}{Runyi Hu}, \bibinfo{person}{Tianwei Zhang}, \bibinfo{person}{Fei Wu}, {et~al\mbox{.}}} \bibinfo{year}{2023}\natexlab{a}.
\newblock \showarticletitle{Instruction tuning for large language models: A survey}.
\newblock \bibinfo{journal}{\emph{arXiv preprint arXiv:2308.10792}} (\bibinfo{year}{2023}).
\newblock


\bibitem[Zhang et~al\mbox{.}(2018)]%
        {zhang2018encoding}
\bibfield{author}{\bibinfo{person}{Yingyi Zhang}, \bibinfo{person}{Jing Li}, \bibinfo{person}{Yan Song}, {and} \bibinfo{person}{Chengzhi Zhang}.} \bibinfo{year}{2018}\natexlab{}.
\newblock \showarticletitle{Encoding conversation context for neural keyphrase extraction from microblog posts}. In \bibinfo{booktitle}{\emph{NAACL}}. \bibinfo{pages}{1676--1686}.
\newblock


\bibitem[Zhao et~al\mbox{.}(2023a)]%
        {zhao2023hierarchical}
\bibfield{author}{\bibinfo{person}{Weihao Zhao}, \bibinfo{person}{Han Wu}, \bibinfo{person}{Weidong He}, \bibinfo{person}{Haoyang Bi}, \bibinfo{person}{Hao Wang}, \bibinfo{person}{Chen Zhu}, \bibinfo{person}{Tong Xu}, {and} \bibinfo{person}{Enhong Chen}.} \bibinfo{year}{2023}\natexlab{a}.
\newblock \showarticletitle{Hierarchical Multi-modal Attention Network for Time-sync Comment Video Recommendation}.
\newblock \bibinfo{journal}{\emph{IEEE Transactions on Circuits and Systems for Video Technology}} (\bibinfo{year}{2023}).
\newblock


\bibitem[Zhao et~al\mbox{.}(2023b)]%
        {zhao2023bootstrapping}
\bibfield{author}{\bibinfo{person}{Xinping Zhao}, \bibinfo{person}{Ying Zhang}, \bibinfo{person}{Qiang Xiao}, \bibinfo{person}{Yuming Ren}, {and} \bibinfo{person}{Yingchun Yang}.} \bibinfo{year}{2023}\natexlab{b}.
\newblock \showarticletitle{Bootstrapping Contrastive Learning Enhanced Music Cold-Start Matching}. In \bibinfo{booktitle}{\emph{WWW}}. \bibinfo{pages}{351--355}.
\newblock


\bibitem[Zheng et~al\mbox{.}(2023)]%
        {zheng2023gpt}
\bibfield{author}{\bibinfo{person}{Shen Zheng}, \bibinfo{person}{Yuyu Zhang}, \bibinfo{person}{Yijie Zhu}, \bibinfo{person}{Chenguang Xi}, \bibinfo{person}{Pengyang Gao}, \bibinfo{person}{Xun Zhou}, {and} \bibinfo{person}{Kevin Chen-Chuan Chang}.} \bibinfo{year}{2023}\natexlab{}.
\newblock \showarticletitle{GPT-Fathom: Benchmarking Large Language Models to Decipher the Evolutionary Path towards GPT-4 and Beyond}.
\newblock \bibinfo{journal}{\emph{arXiv preprint arXiv:2309.16583}} (\bibinfo{year}{2023}).
\newblock


\bibitem[Zhu et~al\mbox{.}(2018)]%
        {zhu2018learning}
\bibfield{author}{\bibinfo{person}{Han Zhu}, \bibinfo{person}{Xiang Li}, \bibinfo{person}{Pengye Zhang}, \bibinfo{person}{Guozheng Li}, \bibinfo{person}{Jie He}, \bibinfo{person}{Han Li}, {and} \bibinfo{person}{Kun Gai}.} \bibinfo{year}{2018}\natexlab{}.
\newblock \showarticletitle{Learning tree-based deep model for recommender systems}. In \bibinfo{booktitle}{\emph{KDD}}. \bibinfo{pages}{1079--1088}.
\newblock


\end{thebibliography}

\end{document}